\documentclass[twocolumn,epjc3]{svjour3}  
\usepackage{amsmath,amssymb,amsfonts,graphicx,color}
\usepackage{macros}
\journalname{Eur. Phys. J. C}
\graphicspath{{./plots/}}

\begin{document}

\title{Controlling the sign problem in Finite Density Quantum Field
  Theory} 

\author{Nicolas Garron\thanksref{e2} \and 
  Kurt Langfeld\thanksref{e1} 
}

\thankstext{e2}{e-mail: nicolas.garron@liverpool.ac.uk}
\thankstext{e1}{e-mail: kurt.langfeld@liverpool.ac.uk}

\institute{\liverpool 
}

\date{Received: date / Accepted: date}

\maketitle

\begin{abstract}
Quantum field theories at finite matter densities generically possess
a partition function that is exponentially suppressed with the volume
compared to that of the phase quenched analogue. The smallness arises
from an almost uniform distribution for the phase of the fermion
determinant. Large cancellations  upon integration is the origin of a
poor signal to noise ratio.  We study three alternatives for this
integration: the Gaussian approximation, the ``telegraphic'' 
approximation, and a novel expansion in 
terms of theory-dependent moments and universal coefficients.
We have tested the methods for QCD at finite densities of
heavy quarks. We find that for two of the approximations 
the results are extremely close - if not identical - to the full answer
in the strong sign problem regime.
\end{abstract}




\section{Introduction}

The sign problem is known to be one the most important challenges of
modern physics. In theoretical particle physics, it prevents us from
simulating finite-density QCD with standard Monte-Carlo methods. Hence
most of the QCD phase diagram cannot be explored by first-principle
techniques, such as lattice QCD. Many reviews can be found, see for
example 
\cite{Fukushima:2010bq,deForcrand:2010ys,Gupta:2011ma,Levkova:2012jd,Aarts:2013lcm,Gattringer:2014nxa,Sexty:2014dxa,Borsanyi:2015axp,Langfeld:2016kty}.

\medskip 
Dropping the phase factor of the quark determinant $\exp \{ i \phi \}$
from the functional integral results in a theory, say with partition
function $Z_{PQ}$, that is accessible by standard importance sampling
Monte-Carlo simulations. Very early on, it became clear that
$Z_{PQ}$ and the partition function of the full theory $Z$ are only
comparable for the smallest values of the chemical potential
$\mu$~\cite{Troyer:2004ge}. The deviation is quantified by the so-called
phase factor expectation value  
\be 
\la \mathrm{e}^{i \phi } \ra _{PQ}  \; = \; Z(\mu ) / Z_{PQ}(\mu ) \;
\propto \; \mathrm{e}^{- \Delta f \, V } \; , 
\en
where $\Delta f$ is the free energy difference between the full and the phase
quenched theory and $V$ is the volume (see
e.g.~\cite{Troyer:2004ge}). 
The knowledge of this phase factor would give access to the partition
function $Z(\mu )$ (we assume that $Z_{PQ}(\mu )$ has been obtained by
standard methods).
In this work, we study its expectation value,
$\la \mathrm{e}^{i \phi } \ra _{PQ} $:
it is a very small number, generically very hard to measure due to the statistical noise, which only
decreases proportionally to the square root of the number of Monte-Carlo configurations.
Our approach is based on the density-of-states method and in particular on the LLR
formulation~\cite{Langfeld:2012ah,Langfeld:2015fua}, which is ideally suited
to calculate probability distributions of observables: it
features an exponential error suppression~\cite{Langfeld:2015fua}
which can result in an unprecedented precision for the observable (see
e.g.~for an early example~\cite{Langfeld:2013xbf}). It is based upon a
non-Markovian Random Walk, which immediately provides two main
advantages: it bears the potential to overcome the critical slowing
down for theories close to a first order phase
transition~\cite{Lucini:2016fid,Langfeld:2016kty}, and it is not
restricted to theories with a positive probabilistic weight for
Monte-Carlo configurations. In fact, the method has been successfully
applied to the $Z_3$ theory at finite
densities~\cite{Langfeld:2014nta} and QCD at finite densities of
heavy quarks~\cite{Garron:2016noc}. In both cases, the
probability density $\rho (\phi )$ of the phase $\phi $ has been
obtained to very high precision. The phase factor expectation value is
then given by 
\be
\label{eq:phase}
\la \mathrm{e}^{i \phi } \ra _{PQ}  \; = \; \frac{ \int d \phi \; \rho
  (\phi ) \; \exp  \{ i \phi \} }{ \int d \phi \; \rho  (\phi )} . 
\en
Despite of high quality numerical result for $\rho (\phi)$, the challenge remains to
extract a very small signal from the above Fourier transform. An 
approach, put forward in~\cite{Langfeld:2014nta,Garron:2016noc},
is to first represent the numerical data for $\ln \rho (\phi )$ by a
fit function and then to calculate the Fourier transform of the fit
function (semi-)analytically. The method produces reliable results if
all the numerical data are well represented by the fit function with a
{\it small}  number of fit parameters~\cite{Langfeld:2014nta,Garron:2016noc}.  
With the advent of high precision data for $\rho (\phi )$, the main
obstacle for gaining access to quantum field theories at finite
densities is the above Fourier transform. The method used
in~\cite{Langfeld:2014nta,Garron:2016noc} hinges on the fact that a
fit function which faithfully represents the data could be found.
This might not be generically the case.

In this paper, we propose three alternatives to this direct method.
In Section~\ref{sec:ga} we present the first approach,
called {\em Gaussian approximation}. No fitting procedure is required,
instead the phase factor is computed directly from the data.
Within this framework, the integral in the numerator of \Eq(\ref{eq:phase})
is known analytically.
The second approximation, presented in Section~\ref{sec:tele} is what
we call the ``telegraphic'' approximation.
This approach can be implemented either on the fit function or
directly on the data (although it might require new simulations).
The integral is replaced by a simple difference.
In Section~\ref{sec:am}, we introduce a third method,
the ``Advanced Moment expansion'', which can be seen as a variant
of a cumulant expansion~\cite{Ejiri:2007ga,Nakagawa:2011eu,Saito:2012nt,Saito:2013vja}.
It is a systematic expansion in the deviation
from the uniform distribution and as such is expected to work better in
the strong sign-problem regime.
We will provide evidence that the universal coefficients decrease exponentially
with increasing order, providing a rapid convergence if the moments are
bounded. Although the convergence is faster in the strong sign-problem regime,
for the phase factor expectation value we find an excellent agreement
already at the third order of the expansion, regardless of the strength
of the sign problem.
In this case we still rely on a fitting procedure for the density of states.
However the direct computation of the Fourier transform \Eq(\ref{eq:phase})
is not needed, only the elementary moments are required. 
%
Before going through the details of these methods, we present the framework
and the numerical details of our simulations in the next section.
Our conclusions are presented in \ref{sec:conclusions}



\section{Generalities and Framework}
\label{sec:2}
\subsection{Full theory and phase quenching}
We consider a generic theory with a partition function
\be
Z \; = \; \int {\cal D }U_\mu  \; \exp \{ \beta \, S_\mathrm{YM}[U] \}\;
\hbox{Det} M[U] \; ,
\label{eq:Z}
\en
and with a complex ``matter'' determinant:
\be
   \hbox{Det} M[U] = |\hbox{Det} M[U] | \;\exp\{i\phi[U] \}\;, \; \; \;
   \phi \in ]-\pi,\pi] \; . 
\en
With the help of the density of states
\be
\rho(s) \; = \; \int {\cal D }U_\mu  \; \exp \{ \beta \,S_\mathrm{YM}[U] \}
|\hbox{Det} M[U]| \delta(s - \phi[U]) \;, 
\en
the partition function can then be recovered by a 1-dimensional
Fourier transform: 
\be
Z \; = \; \int ds \; \rho(s) \; \exp \{ i  s\}   \;
= \; \int ds \; \rho(s) \; \cos(s)  \;. 
\en
We also introduce the so-called  {\em phase quenched}
counter part by 
\bea
\label{eq:Zpq}
Z_{PQ} & = & \int {\cal D }U_\mu  \; \exp \{ \beta \,
S_\mathrm{YM}[U] \}\; \vert \hbox{Det} M[U] \vert  \;, \\
&=&
\; \int ds \; \rho(s) \;. 
\ena
The expectation values of an observable $A$ in the full and in the
phase quenched theory are given as usual by
\bea
\la A \ra     &=& \frac{1}{Z} \int {\cal D }U_\mu  \; A \exp \{ \beta
\, S_\mathrm{YM}[U] \} \hbox{Det} M[U] \;, \\
\la A \ra_{PQ} &=& \frac{1}{Z_{PQ}} \int {\cal D }U_\mu  \; A \exp \{
\beta \, S_\mathrm{YM}[U] \} |\hbox{Det} M[U]| \;, \nn\\
\ena
implying the well-known relations
\be
\la A \ra =  \frac{ \la A \e^{i \phi}\ra_{PQ}}{ \la \e^{i
    \phi}\ra_{PQ}} \;  \hbo 
Z \; = \; Z_{PQ} \; \la \e^{i\phi} \ra_{PQ} \; . 
\en
In terms of the density, the phase factor expectation value is given by
\Eq(\ref{eq:phase}).

\subsection{Extensive density of states}

For theories for which the imaginary part arises from a local action
an extensive phase $x \in ]-\infty, \infty [$ can be defined as
the sum of the local phases. This has been e.g. the case for the
finite density $Z_3$ and for heavy dense
QCD~\cite{Langfeld:2014nta,Garron:2016noc}. For fermionic theories
with the phase $\phi [U]$ arising from the (non-local) quark determinant, an
extensive phase can still be defined as pointed out
in~\cite{Greensite:2013gya}:
\bea
x[U] &=& \hbox{Im ln} \, (\Det M)
\nonumber \\
&=& \int _0^{\mu/T} \hbox{Im} \, \left[ \frac{\partial ( \hbox{ln} \,
    Det) M) }{ \partial     \mu/T } \right]_{\mu=\bar{\mu}} \; d \left(
\frac{\bar{\mu}}{T} \right)
\nonumber \\
&=&  \int _0^{\mu/T} \hbox{Im} \, \hbox{tr} \left[ M^{-1} \frac{\partial M)
  }{ \partial     \mu/T } \right]_{\mu=\bar{\mu}} \; d \left( 
\frac{\bar{\mu}}{T} \right)
\ena 
The definition of an {\it extensive} phase factor has proven to be
important to achieve the precision needed for the Fourier
transform. If $\rho _E(x) $ denotes the corresponding probability
distribution, the phase factor expectation value in \Eq(\ref{eq:phase}) is
obtained by 
\be
\label{eq:phase1}
\la \e^{i\phi} \ra_{PQ}  = \frac{1}{Z_{PQ}}
\int_{-\infty}^{+\infty} dx \; \rho _E (x) \cos(x)  \;.
\en
The density of states $\rho (s)$ can be easily recovered from the extended
density $\rho _E(x)$. To see this, we subtract from $x$ a multiple of
$2 \pi$ until $s \in [-\pi,\pi[$, $s=x-2\pi \, n$, $n \in \mathbb{Z}$,
and split the integration domain in intervals of size $2\pi$:  
\bea
\label{eq:phase_rhoF}
\la \e^{i\phi} \ra_{PQ}
&=& \frac{1}{Z_{PQ}} \sum_{n\in {\mathbb Z}} \int_{-\pi }^{\pi} ds \;
\rho _E(s+2n\pi) \cos(s) \;, \nn \\
&=& \frac{1}{Z_{PQ}} \int_{-\pi }^{\pi} ds \, \rho (s) \, \cos(s) \;,
\ena
and identify: 
\be
\rho (s)  \equiv  \sum_{n\in {\mathbb Z}} \rho _E (s+2\pi n) \; .
\label{eq:re}
\en
Also note that 
$$ 
Z_{PQ} \; = \; \int _{-\pi}^\pi ds \; \rho (s) \; = \; \int_{-\infty
}^\infty dx \; \rho_E(x) \; . 
$$

\subsection{Volume dependence of the density} 

We here consider the class of theories for which the phase of the
Gibbs factor is proportional to the chemical potential $\mu $ and for
which this is the only $\mu $ dependence. Scalar theories do not fall
into this class since the real part of the action also acquires a
$\mu $ dependence, but fermion theories in the ab initio continuum
formulation might fall into this class. 
For these theories, let us study the dependence of $\rho_E (s) $ on the
physical volume $V$. We make explicit the $\mu $ dependence of the
phase factor expectation value and point out that the partition function
is positive for all $\mu $: 
\be
z(\mu ) \; = \; \la \e^{i \, \mu \, \phi } \ra _{PQ} \; \geq \; 0 \; . 
\label{eq:z1}
\en
Note that we have $z(0)=1$ and that we will assume that 
\be 
\la \e^{- i \, \mu \, \phi} \ra_{PQ} \; = \; \la \e^{i \, \mu \, \phi}
\ra_{PQ} \; \; \Rightarrow \; \; z(-\mu)=z(\mu) \; . 
\label{eq:01b} 
\en
Note that since $z(\mu)$ is obtained by a Fourier transform of $\rho$,
see \Eq(\ref{eq:phase}), the density of states can be recovered from
$z(\mu )$ by the inverse Fourier transform
(up to a normalisation constant $Z_{PQ}\ge 0$)
\be
\rho_E (s) \; = \; Z_{PQ} \; \int
\frac{d\mu}{2\pi} \; z(\mu) \, \e ^{-i s \mu } \; . 
\label{eq:z2}
\en
As argued in~\cite{Greensite:2013gya}, 
$z(\mu )$ can be viewed as a partition function with free energy
density $f(\mu)$
(a necessary condition is that $z(\mu) \ge 0 $), 
leaving us with the volume dependence:
\bea 
z(\mu) &=& \exp \{ - f(\mu) \, V \} \;,\\
&=& \exp \{ - [c_1 \mu^2 + c_2 \mu ^4 + c_3 \mu ^6 + \ldots ] \; V \} \; ,  
\label{eq:z3}
\ena
where the coefficients $c_k$ are volume independent. 
Inserting \Eq(\ref{eq:z3}) into \Eq(\ref{eq:z2}), we find with an expansion
in inverse powers of $V$:
\be
\rho_E (s) = \mathrm{const.} \; \exp \left\{
- a_1 \, \frac{ s^2 }{V} - a_2 \, \frac{ s^4 }{V^3 } - a_3 \, \frac{
  s^6 }{V^5 } + \ldots \right\}  \; ,
\en
\bea 
a_1 &=&\frac{1}{4 c_1} \; + \; {\cal O}(1/V) , \; \; \;
a_2 \; = \; \frac{ c_2 }{16 c_1^4} \; + \; {\cal O}(1/V)  
\nonumber \\
a_3 &=& - \frac{ 3 c_2^2 + c_3 c_2 }{c_1^7}  \; + \; {\cal O}(1/V). 
\nonumber 
\ena
If we define a ``scaling'' variable by $x=s/\sqrt{V}$, the deviation
from a Gaussian distribution decreases with increasing volume: 
\be
\rho_E (s) = \mathrm{const.} \; \exp \left\{
- a_1 \, x^2 - a_2 \, \frac{ x^4 }{V} - a_3 \, \frac{
  x^6 }{V^2 } + \ldots \right\}  \;  . 
\label{eq:z4}
\en

\subsection{Numerical details}
We use the data obtained in our previous work~\cite{Garron:2016noc}
but have also generated new simulations for reasons that we explain
below.
We summarise here the parameters used for the numerical simulations
and the methods to obtain the density of states.
The interested reader will find more details in the aforementioned reference.
The lattice parameters are 
$$
\hbox{$8^4$ lattice} , \; \; \beta = 5.8, \; \; \kappa = 0.12 \; . 
$$
and we let the chemical potential $\mu$ vary between 1.0421
and 1.4321. We identified the ``strong sign problem region''
as being $1.1 < \mu < 1.4$.
We for each value of $\mu$,
we split the domain of the phase $s \in\left[0,s_{\rm max} \right]$
in $n_{\rm int}$ small interval of size $\delta_s$ 
and on each interval $k$, we compute the LLR coefficients $a_k$.
In practise we choose $s_{\rm max} \sim 36$, 
$\delta_s =0.896$ and $n_{\rm int} = 40$,
except for a few values of the chemical potential,
for which we need a better resolution. The corresponding
values are reported in \Table{\ref{table:deltas}}.
\begin{table}[htb]
  \begin{center}
   \begin{tabular}{c| l l }
     \hline \\[-2ex]
     $\mu$ & $\delta_s$ & $n_{\rm nint}$ \\
     \hline
 1.1821  &  0.29867 & 120\\
 1.3721  &  0.4480  & 80\\
 1.3921  &  0.4480  & 80\\
 1.4121  &  0.4480  & 80\\
 \hline
   \end{tabular}
   \caption{Size and number of intervals for the LLR simulations. 
   For the other values of $\mu$, we choose $\delta_s =0.896$ and $n_{\rm int} = 40$.}
   \label{table:deltas}
   \end{center}
\end{table}
We reconstruct the probability density function for discrete values
of the phase $s_k=k \delta_s +\delta_s /2 $, namely
\be
\rho_E (s_k) = \exp \left\{ -\sum_{i=1}^{k-1} a_i \delta_s
- a_k \delta_s/2\right\} \;. 
\label{eq:rho_recons}
\en
In~\cite{Garron:2016noc} we performed a polynomial fit of $\ln(\rho_E)$
and computed \Eq(\ref{eq:phase1}) by a semi-analytic integration
(we refer to this method as ``Exact'').

\begin{figure}[t]
  \begin{center}
  \includegraphics[width=0.5\textwidth]{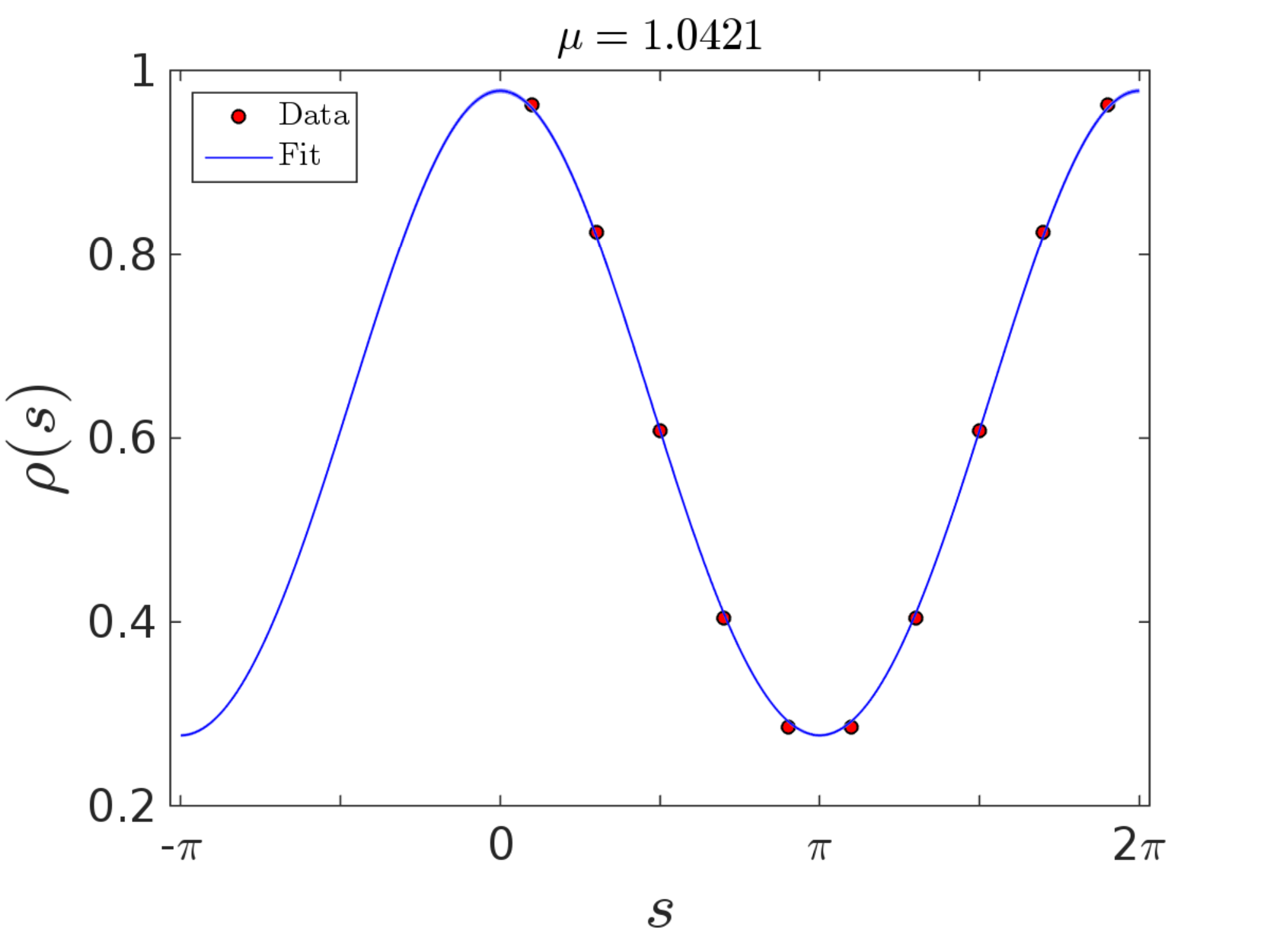}
  \caption{The density obtained directly from the data
    or from fitting the extensive density $\rho _E$, in the low
    density-region   where the sign problem is weak.}
  \label{fig:1}
    \end{center}
\end{figure}

\begin{figure}[t]
  \begin{center}
      \includegraphics[width=0.5\textwidth]{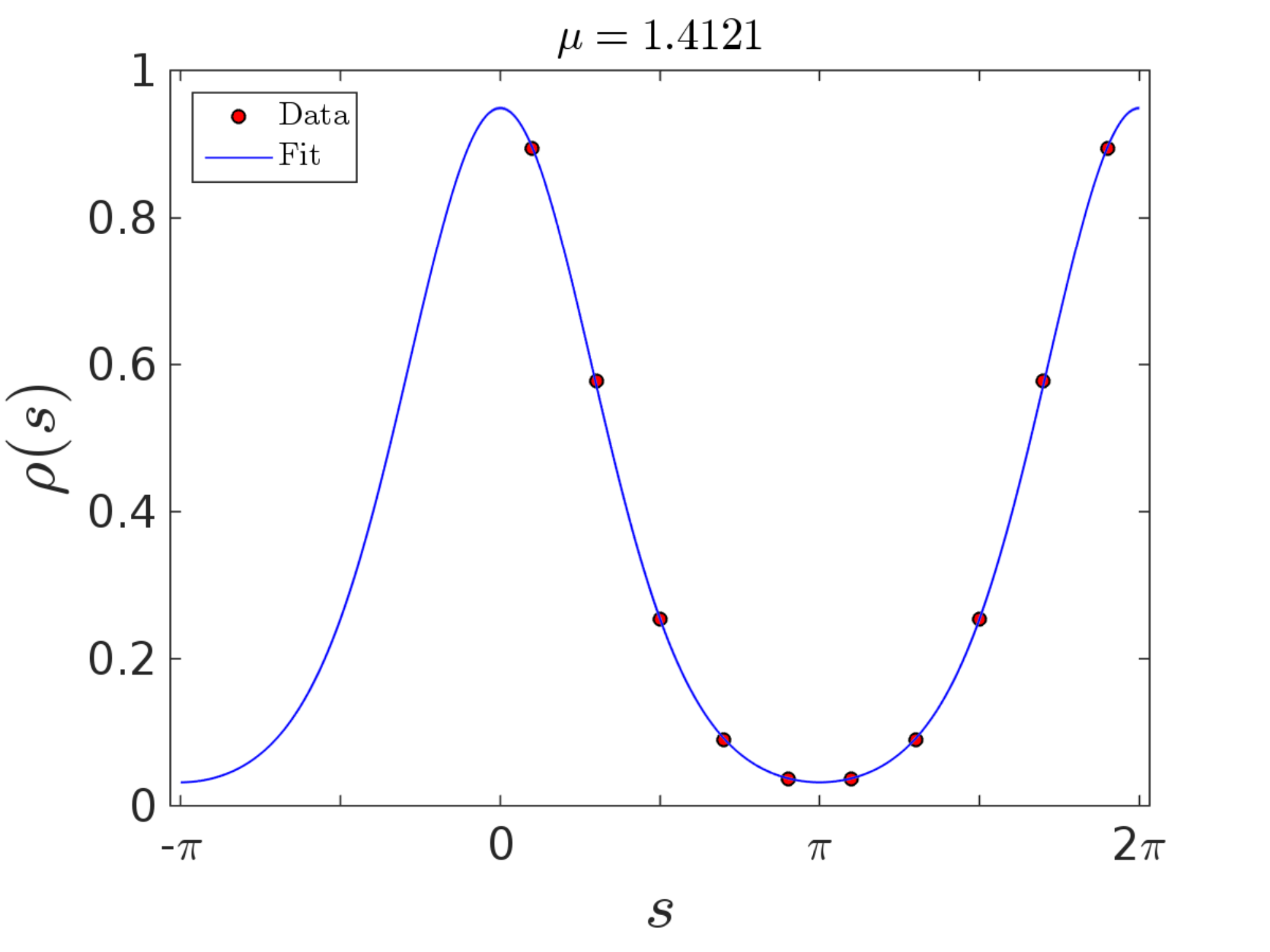} 
  \caption{Same as Fig.~\ref{fig:2} but with the density close to
    offset, again the sign problem is weak} 
  \label{fig:2} 
    \end{center}
\end{figure}

\begin{figure}[t]
  \begin{center}
  \includegraphics[width=0.5\textwidth]{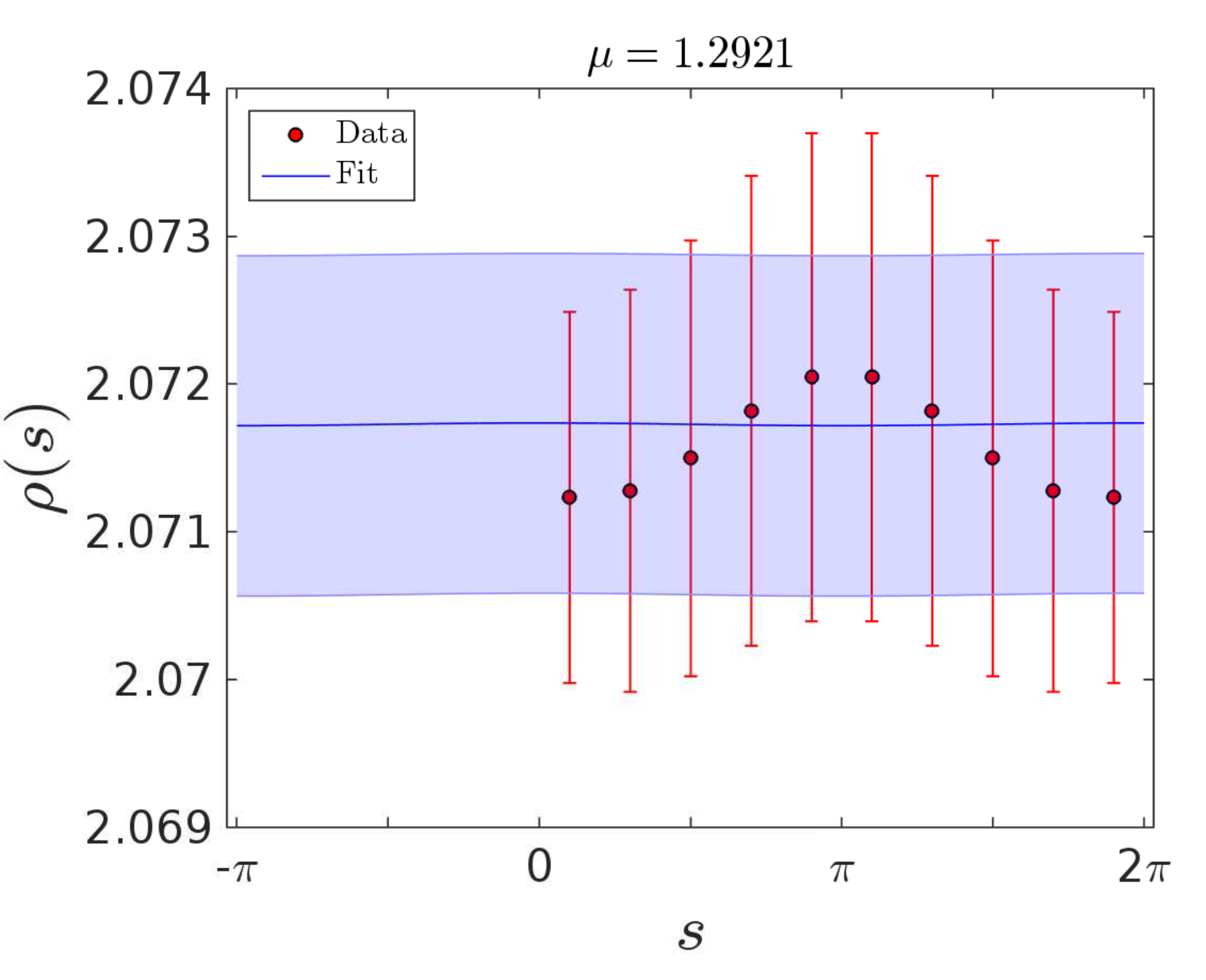}
  \caption{Same as previous figures but in the strong sign problem
    regime. The blue band   corresponds to the $1-\sigma$ region
    obtained from the fit.} 
  \label{fig:3}
  \end{center}
\end{figure}

Although the fits are of very good quality and very stable,
for three values of the chemical potential, we have also ran new simulations
with $\delta_s = \pi/5$.
As shown below, these new data allow us to compute $\rho(s)$ directly from the data
(without relying on any fitting procedure) and will be very useful
to check the methods presented here.
We have implemented this technique for three
different values of the chemical potential. This is illustrated in
\Figs{\ref{fig:1},\ref{fig:2}} and \ref{fig:3}, 
where we see that the different methods give compatible results.

Finally, we mention that we use around $1000$ configurations
and that the statistical errors are estimated with the bootstrap
method, using 500 samples. Naturally we have checked that the errors
are stable with respect to the number of samples.

\section{The Gaussian approximation \label{sec:ga} }
\label{sec:3}
\begin{figure}[t]
  \begin{center}
    \includegraphics[width=0.4\textwidth]{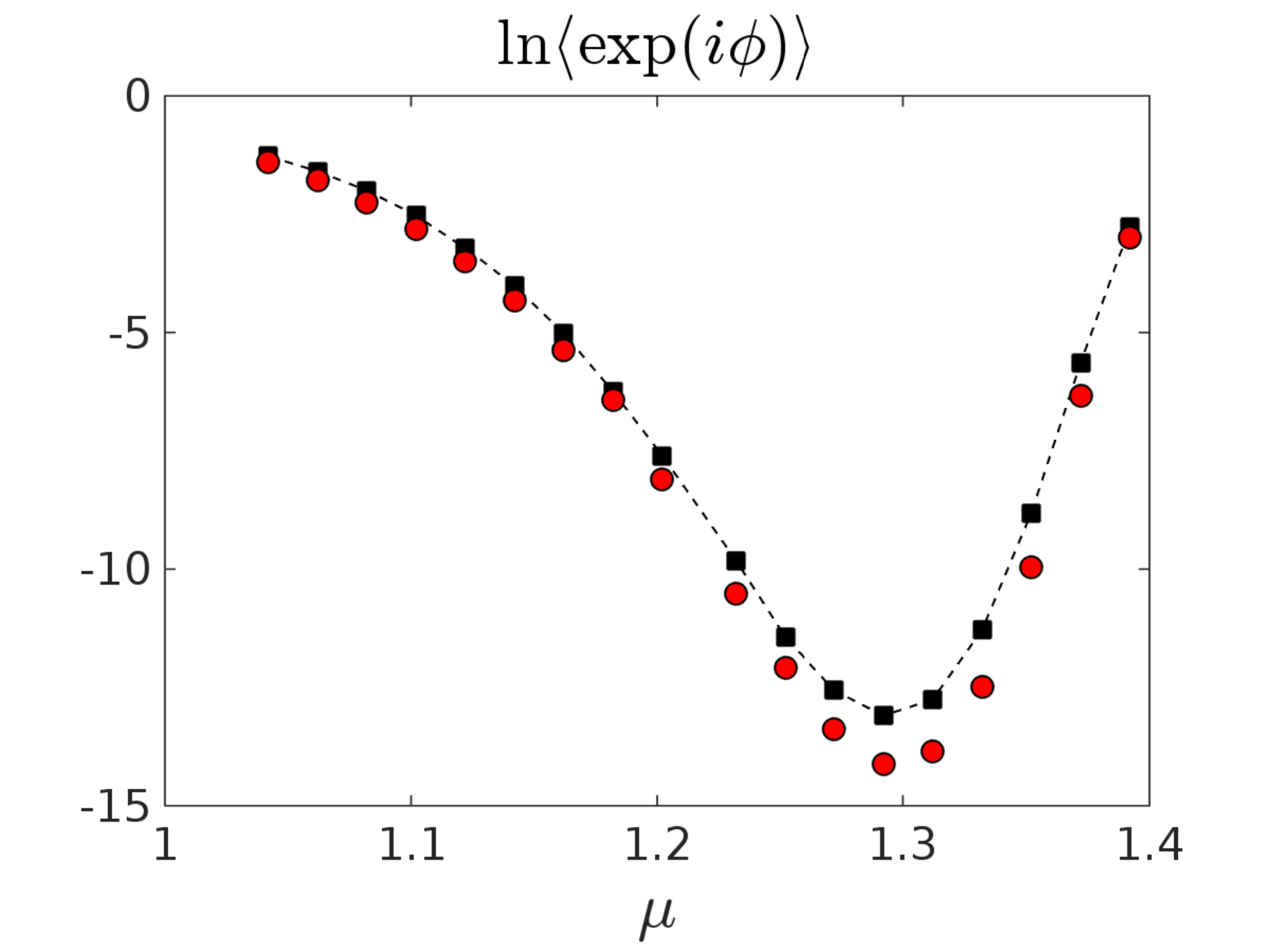}
    \caption{ The phase factor expectation value in Gaussian
      approximation (red symbols) in comparison with the exact result
      (black) from~\cite{Garron:2016noc}.
    }
  \label{fig:c} 
    \end{center}
\end{figure}

The smallness of $\la \mathrm{e}^{i \phi } \ra_{PQ} $
arises from large cancellations in~\Eq(\ref{eq:phase}).
It was pointed out be Ejiri~\cite{Ejiri:2007ga} that these
cancellations can be avoided by using cumulants of the phase
factor: 
\be
\la \mathrm{e}^{i \phi } \ra _{PQ}  \; = \; \exp \left[ - \frac{1}{2}
  \la \phi^2 \ra _c \, + \, \frac{1}{4!} \, \la \phi ^4 \ra _c - \,
  \ldots \right] \; . 
\en
In fact, numerical results suggest that the probability distribution
is Gaussian to a good
extent~\cite{Ejiri:2007ga,Ejiri:2012ng,Ejiri:2012wp,Ejiri:2013lia},
which would imply that only the cumulant $\la \phi^2 \ra _c $ is
non-vanishing. It has been argued in~\cite{Greensite:2013gya} that
higher cumulants are suppressed by factors of the volume $V$ and that,
however, higher order cumulants are important for the medium and high
range of chemical potentials.
Throughout this paper, we define the {\it Gaussian approximation} as
the approximation of the extended density of states by a normal
distribution: 
\be
\rho_E (s) \approx \mathrm{const.} \; \exp \left\{
- \epsilon  \, s^2 \right\} \; .
\label{eq:ga1}
\en
The phase factor expectation value \Eq(\ref{eq:phase}) is then
analytically obtained:
\be
\la \mathrm{e}^{i \phi } \ra _{PQ}  \; = \;
\exp \left\{ - \frac{1}{4 \epsilon } \right\} \; .
\label{eq:ga2}
\en
We extract the parameter $\epsilon$ from the standard expectation
value by 
\be
\label{eq:s2gauss}
\langle s^2 \rangle_E \; = \; \frac{1}{2 \, \epsilon } \; \; \;
\Rightarrow \; \; \; \la \mathrm{e}^{i \phi } \ra _{PQ}  \; = \;
\exp \left\{ - \frac{1}{2 } \langle s^2 \rangle_E \right\} ,  
\en
where the subscript $E$ indicates that the expectation values are
defined with respect to the extended density $\rho _E$. 
We test this approach for heavy-dense QCD with partition function
(\ref{eq:Zpq}).
We find the expectation value in \Eq~(\ref{eq:s2gauss}) directly
from the data:
we take the density obtained through \Eq(\ref{eq:rho_recons})
and compute the  expectation value $\langle s^2 \rangle_E$ using a trapezoidal
approximation. We obtain  in this way an estimate for the phase factor
expectation value \Eq(\ref{eq:ga2}) without invoking any fitting procedure.
Our numerical findings are summarised in \Fig{\ref{fig:c}}. We find
that the Gaussian approximation provides a surprisingly good
approximation over the whole range of chemical potentials $\mu $. Even
in the strong sign-problem regime at intermediate values $\mu $, the
cancellations are well emulated and the approximate result only 
underestimates the true result by roughly a factor $2$. 

\section{The ``telegraphic'' approximation}
\label{sec:tele}
\subsection{Methodology}
As can be seen in \Fig{\ref{fig:3}}, $\rho$ weakly depends on
its arguments in the strong sign-problem regime
and for large volumes.
In this case, a Poisson re-summation of (\ref{eq:re}) should
yield a rapidly converging series:
\bea
\rho (s) &=& \sum _{\nu \in \mathbb Z} b_\nu
\label{eq:pr} \;,\\
b_\nu &=& \int _{-\infty} ^\infty dn \; \e ^{2\pi i \, \nu \, n} \, \rho
_E (s+2\pi n)
\nonumber \\
&=& \frac{1}{2\pi }  \; \e ^{-i\,  \nu \, s} \; \int _{-\infty} ^\infty
dx \;  \e ^{i\,  \nu \, x} \; \rho_E(x) \; .
\nonumber
\ena 
The sum over $\nu $ in (\ref{eq:pr}) becomes: 
\bea
\rho (s) &=& c_0 \; + \; \sum_{\nu=1}^\infty c_\nu \; \cos (\nu \, s)
\; ,
\label{eq:pr2} \\
c_0 &=&  \frac{1}{2\pi } \int _{-\infty} ^\infty
dx \;   \rho_E(x) \; , 
\label{eq:pr3} \\
c_\nu &=&  \frac{1}{\pi } \int _{-\infty} ^\infty
dx \;  \cos(\nu x) \; \rho_E(x) \; , \; \; \; \nu \ge 1.
\ena
Note that we find in view of (\ref{eq:phase1})
\be 
c_1/c_0 \; = \; 2 \; \la \e^{i\phi} \ra_{PQ} \; . 
\label{eq:la1}
\en
If the sum over $\nu $ is rapidly converging, we find approximately: 
\be
\rho (s) /c_0 \; \approx \; 1 \; + \; 2 \; \la \e^{i\phi} \ra_{PQ} \;
\cos (s) \; . 
\label{eq:pr4}
\en 
In the strong sign-problem regime, the amplitude of the cosine is very
small, and therefore we see that $\rho (s)$ is almost a constant. Equation
(\ref{eq:pr4}) then offers the possibility to extract the phase factor
expectation value, i.e.,
\be
\la \e^{i\phi} \ra_{PQ} \; \approx \; \frac{1}{4c_0} \Bigl[
  \rho (0) - \rho (\pi) \Bigr] \; .
\label{eq:pr5}
\en 
Using (\ref{eq:re}), we therefore find:
\be
\la \e^{i\phi} \ra_{PQ} \; \approx \; \frac{\pi}{2} \;
\frac{ \sum _{k\in \mathbb Z} (-1)^k \, \rho _E ( k \pi) }{
  \int _{-\infty} ^\infty dx \; \rho_E(x) }  \; . 
\label{eq:pr6}
\en 
We call this the {\it telegraphic approximation}. It emerges by
neglecting higher contributions $c_\nu $ of the Poisson sum. In order
to get a feeling for the resulting systematic error, we adopt, for now
only, the Gaussian approximation (\ref{eq:ga1}) and find:
$$
\frac{c_2}{c_1} \; \approx \; \left[ \exp \left(- \frac{1}{4 \epsilon}
  \right) \right]^3 . 
$$
This implies that the correction to $\rho (s) $ in (\ref{eq:pr4}) is
of order:
$$
\frac{c_2}{c_0} = \frac{c_2}{c_1}\frac{c_1}{c_0} \; = \; \approx \;
2 \, \left[ \la \e^{i\phi} \ra_{PQ} \right] ^4 \; , 
$$
where we have used (\ref{eq:ga2}) and (\ref{eq:la1}). At least in the
strong sign-problem regime, for which $\la \e^{i\phi} \ra_{PQ}$ is
very small, we expect the telegraphic approximation to work very
well. 

\medskip
We finally point out that the telegraphic approximation can be
improved in a systematic way. The order of the approximation is
defined by the number of harmonics entering the density of
states. E.g., in 3rd order we have: 
\bea
\rho (s) /c_0 & \approx & 1 \; + \; 2 \; \la \e^{i\phi} \ra_{PQ} \;
\cos (s) \;  + \; c \; \cos (2s) 
\label{eq:pr4b} \\ 
&+&  d \; \cos (3s) \; , 
\nonumber 
\ena 
with the unknowns $\la \e^{i\phi} \ra_{PQ} $ and $c,d$. We generate
three equations by evaluating $\rho (s)$ at $s=0,\,  \pi/3, \, \pi $
and solve the linear set of equations for the unknowns. We are
predominantly interested in the phase factor: 
\bea 
\la \e^{i\phi} \ra_{PQ}  &\approx & 
- \frac{1}{2} \; + \; \frac{1}{4} \rho (0) /c_0  \; + \; \frac{1}{3}
\rho (\pi/3) /c_0 
\label{eq:pr4c} \\ 
&-&   \frac{1}{12} \; \rho (\pi) /c_0 \; , 
\nonumber 
\ena 
which can be easily converted to a discrete sum over discrete set of
points of $\rho _E(s)$ using (\ref{eq:re})\; .

\subsection{Numerical implementation}
\begin{figure}[t]
  \begin{center}
    \includegraphics[width=0.4\textwidth]{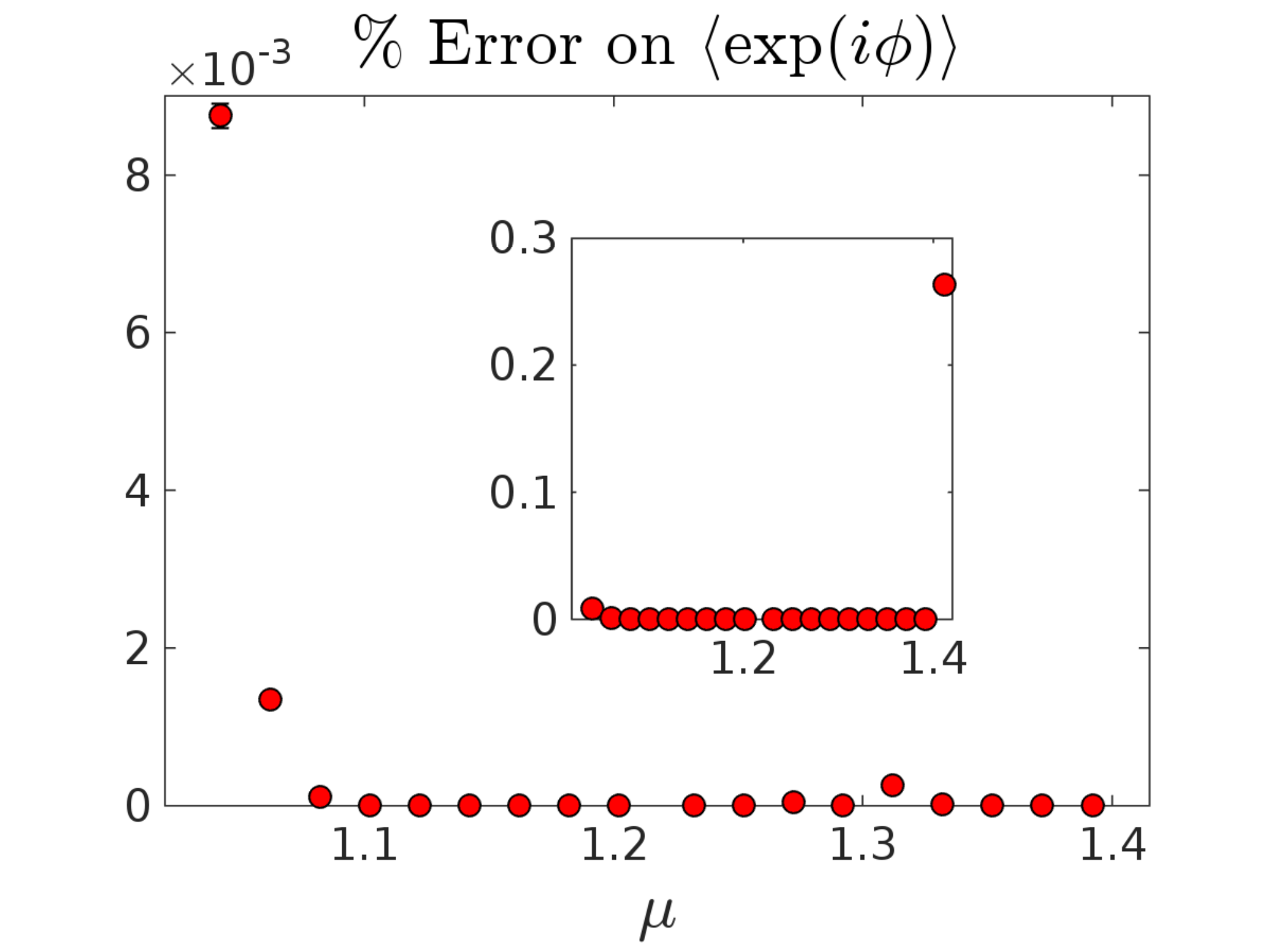}
    \caption{Relative difference (in percentage) on the phase factor expectation value between
      the telegraphic approximation
      \Eq(\ref {eq:pr6}) and the ``exact'' answer~\cite{Garron:2016noc}.
      Not visible in the plot is the point
      $\mu=1.4121$, for which the relative error is $0.3\%$}
  \label{fig:firstapprox} 
    \end{center}
\end{figure}

Again, we use Heavy-Dense QCD to test this approximation.
Having in hands the density of state - 
either $\rho_E$ obtained from the fit or $\rho$ from the date through \Eq(\ref{eq:re})
- it is straightforward to implement numerically \Eq(\ref {eq:pr6}). 
If we take the results from the fit, we find that this approximation provide
results extremely close to the ``exact'' ones: except for a few values of $\mu$
in the weak sign problem regime, the results (central value and variance)
are actually indistinguishable. For example, for $\mu = 1.0821$, we find
\bea
\ln \la \e^{i\phi} \ra_{PQ}^{\rm exact} &=& -1.992175 \pm  2.910279
\times 10^{-3}  \;,\\  
\ln \la \e^{i\phi} \ra_{PQ}^{\rm approx} &=& -1.992174 \pm 2.910306
\times 10^{-3} \;. 
\ena
We show our results for the various $\mu$ 
in \Table{\ref{table:resultsmu_approx}} and \Fig{\ref{fig:firstapprox}}.

We have also implemented this approximation for our new simulations where $\delta_s = \pi/5$, such that
we can compute $\rho(s)$ directly from the data
(without relying on any fitting procedure).
In that case we have $\rho(s)$ for $s=\pi/10, 3\pi/10, \ldots$,
but do not have  $\rho(0)$ nor $\rho(\pi)$.
Therefore we use a variant of \Eq(\ref{eq:pr5}):
\be
\la \e^{i\phi} \ra_{PQ} \; \approx \; \frac{1}{4c_0 \cos(\delta_s/2)} \Bigl[
  \rho (\delta_s/2) - \rho (\pi-\delta_s/2) \Bigr] \; .
\label{eq:pr5b}
\en 
In that case we find
\bea
\mu = 1.0421\;, \qquad  &\la \e^{i\phi} \ra_{PQ}^{\rm approx}& = 0.2882(8) \;,   \\
\mu = 1.2921\;  \qquad  &\la \e^{i\phi} \ra_{PQ}^{\rm approx}& = -0.0010(10) \;, \\
\mu = 1.4121\;, \qquad  &\la \e^{i\phi} \ra_{PQ}^{\rm approx}& = 0.6094(18)\;.
\ena
Using the same data, the ``exact'' results obtained through the fit yield
\bea
\mu = 1.0421\;, \qquad  &\la \e^{i\phi} \ra_{PQ}^{\rm exact}& = 0.2838(5) \,,\\ 
\mu = 1.2921\;, \qquad  &\la \e^{i\phi} \ra_{PQ}^{\rm exact}& = 2.19(32) \times 10^{-6} \;,\\
\mu = 1.4121\;, \qquad  &\la \e^{i\phi} \ra_{PQ}^{\rm exact}& = 0.5941(15) \;.
\ena 
Although in the strong sign-problem regime $\mu=1.2921$, we could not extract a signal only 
from the data, for the two other values of $\mu$ we find a decent agreement.

\begin{table}[htb]
  \begin{center}
  \begin{tabular}{c| c c }
    \hline \\[-2ex]
    $\mu$   &  Exact        & $\Delta = $ Exact-Approx    \\
    \hline
    1.0421  & -1.2788(23)   & $-8.746(154)\times 10^{-5}$  \\
    1.0621  & -1.5889(31)   & $-1.350(33) \times 10^{-5}$  \\
    1.0821  & -1.9922(29)   & $-1.056(30) \times 10^{-6}$  \\
    1.1021  & -2.5177(35)   & $-2.258(117)\times 10^{-8}$  \\
    1.1221  & -3.2130(51)   & $1.952(193) \times 10^{-11}$ \\
    1.1421  & -4.0199(92)   & $-2.46(130) \times 10^{-12}$ \\ 
    1.1621  & -5.0194(94)   & $-6.93(113) \times 10^{-14}$ \\
    1.1821  & -6.2506(86)   & $-1.147(13) \times 10^{-12}$ \\
    1.2021  & -7.6034(265)  & $-5.361(148)\times 10^{-9}$  \\
    1.2321  & -9.8246(605)  & $-3.611(224)\times 10^{-11}$ \\
    1.2521  &-11.4458(583)  & $-1.008(66) \times 10^{-8}$  \\
    1.2721  &-12.5563(680)  & $-4.793(355)\times 10^{-7}$  \\
    1.2921  &-13.0923(729)  & $ 1.563(152)\times 10^{-8}$  \\
    1.3121  &-12.7537(1024) & $-2.586(288)\times 10^{-6}$  \\
    1.3321  &-11.2881(493)  & $-1.197(67) \times 10^{-7}$  \\
    1.3521  & -8.8120(156)  & $-2.393(53) \times 10^{-10}$ \\
    1.3721  & -5.6369(203)  & $-9.59(241) \times 10^{-14}$ \\
    1.3921  & -2.7540(92)   & $-9.204(522)\times 10^{-8}$  \\
    1.4121  & -0.83152(297) & $-2.627(49)  \times 10^{-3}$ \\
    \hline
  \end{tabular}        
  \caption{Logarithm of the phase factor expectation value from ~\cite{Garron:2016noc}
    and comparison with the telegraphic approximation presented in the text.
    ($\Delta$ is the deviation for the logarithm of the phase,
    $\Delta = \ln \la \e^{i\phi} \ra_{\rm PQ}^{\rm exact} - \ln \la \e^{i\phi}\ra_{\rm PQ}^{\rm approx}$).}
  \label{table:resultsmu_approx}
\end{center}
\end{table}

\section{The advanced moments approach \label{sec:am}}
\subsection{General formulation } 
The starting point is the expansion of the density-of-states: 
\be
\rho (s) =  \sum_{j=0}^{N_0-1} d_j s^{2j}  \; . 
\label{eq:expansion}
\en
The coefficients
$d_{j }$ depend on the underlying theory, and $N_0 \ge 2$ will define
the order of the expansion. Our conjecture is that the coefficients
$d_j$ are suppressed by powers of the volume with increasing $j$. For
QCD, this conjecture is supported by the strong coupling expansion and
the hadron resonance gas model~\cite{Greensite:2013gya}. There is also some 
numerical evidence by the WHOT-QCD
collaboration~\cite{Ejiri:2012ng,Ejiri:2012wp,Ejiri:2013lia}. Last but not
least, this conjecture becomes true for the limited class of theories
considered in subsection 1.2. Using \Eq(\ref{eq:expansion}) in
\Eq(\ref{eq:phase_rhoF}), we can express the phase factor expectation in
terms of the theory-dependent coefficients $d_j$: 
\be
\la  \e^{i\phi}\ra_{PQ}  = \frac{1}{Z_{PQ}} \sum_{j=1}^{N_0-1} d_j \, I_{2j} \;,
\label{eq:phase_exp1}
\en
where $d_0$ has dropped out upon integration,  and where 
\be
I_{2j} = \int _{-\pi}^\pi ds \, s^{2j} \,\cos (s) 
= \sum_{l=1}^{j}   (-1)^{j-l+1}  \, \frac{2(2j)!}{(2l-1))!} \, \pi^{2l-1}\;.
\label{eq:I2k}
\en
The values $I_{2k}$ can be efficiently calculated by the recursion 
\be
I_{2k} = -2(2k)\pi^{2k-1} - (2k)(2k-1)I_{2k-2} \;,
\en
with the initial condition $I_0=0$. 
Our strategy to access the coefficients $d_j$ in an actual numerical
simulation is to calculate combinations as the simple moments 
$\la s^{2n} \ra $. Using the truncation \Eq(\ref{eq:expansion}) for a
given $N_0$, we find: 
\be 
\langle s^{2n + 2}\rangle \; = \;  \frac{1}{Z_{PQ}} \, \sum_{j=0}^{N_0-1} 
A_{nj}  \; d_j 
\label{eq:k1}
\en
with 
\be
A_{ij} = \frac{2\pi^{2i +2j+1}}{2i +2j+1} \; . 
\label{eq:Aij}
\en 
Keeping in mind that we have $\la s^{2n+2} \ra $ available from a
numerical simulation, the idea is to choose a set of $n$-values and to
consider \Eq(\ref{eq:k1}) as a linear set of equations to obtain the unknowns
$d_j$. Note that for $n=-1$, $\la s^{2n+2}\ra = \la 1 \ra = 0$ follows
from the symmetry $\rho (-s) = \rho (s)$ 
and does not contain theory specific information. We hence choose $n= 0,  \ldots, N_0-1 $
and obtain 
\be
\frac{ d_j }{ Z_{PQ} } \; = \; \sum _{n=0}^{N_0-1} \left( A^{-1}
\right)_{j n}  \, \la s^{2n+2} \ra \; . 
\en
Inserting this into \Eq(\ref{eq:phase_exp1}), we obtain:
\bea
\la  \e^{i\phi} \ra_{PQ}  &=&\sum_{n=0}^{N_0 - 1} \, 
k_n ^{(N_0-1)} \, \la s^{2n+2} \ra \; , \; \; \; N_0 \geq 2 \; , 
\label{eq:k2} \\ 
k_n ^{(N_0-1)} &=& \sum_{j=0}^{N_0-1} \, 
I_{2j}  \; \left( A^{-1} \right)_{j n}  \; , \; \; \; n=0, \ldots, N_0-1
\; .
\label{eq:k2b} 
\ena 
We now have at our fingertips the moment
expansion of the phase factor for a given order $N_0$. We have not yet
achieved a systematic expansion, featuring increments of decreasing
size (when we increase the order $N_0$). To this aim, we define
the first {\it advanced moment} $M_4$ for $N_0=2$ by
\bea 
M_4 &=& \la s^4 \ra \, + \, (k^{(1)}_0 / k^{(1)}_1) \, \la s^2 \ra \; , 
\label{eq:k3} \\ 
\alpha _{4} &=& k^{(1)}_1 \;,
\label{eq:k4} 
\ena
such that, at leading order: 
\be
\la  \e^{i\phi} \ra_{PQ} \; = \; \alpha _4 \, M_4 \; , \; \; \;
N_0 = 2 \;. 
\en
We then define recursively for $N=2, \ldots, (N_0-1)$: 
\bea 
M_{2N +2} &=&  \la s^{2N+2} \ra \nn\\
&+& \frac{ 1 }{k_{N} ^{(N)}} 
\Biggl[ \sum_{n=0}^{N-1} \, k_n ^{(N)}\, \la s^{2n+2} \ra 
- \sum _{n=1}^{N-1} \alpha _{2n+2} \, M_{2n+2} \Biggr] \; , 
\nn\\
\label{eq:k5} \\ 
\alpha _{2N+2 } &=& k_{N} ^{(N)} \; . 
\label{eq:k6} 
\ena 
and finally achieve the systematic expansion:
\be 
\la  \e^{i\phi} \ra_{PQ} \; = \; \sum _{n=1}^{N_0-1} \alpha _{2n+2} \, 
M_{2n+2} \; . 
\label{eq:k7}
\en 
We stress that the coefficients $\alpha_{2n+2}$ are universal, i.e., 
the only dependence on the theory under investigations enters via the
moments $M_k$. Last but not least, we would like to have an explicit
representation of the advanced moments $M$ in terms of the simple
expectations values $\la s^n \ra$. We define: 
\be 
M_{2k} \; = \; \sum _{i=1}^k \gamma _{ki} \, \la s^{2i} \ra \; . 
\label{eq:k8}
\en 
By construction of the advance moments, we have the normalisation 
$\gamma _{kk} =1$. Although for high order 
$N \gg 1$
the intermediate coefficients $\gamma _{Ni}$ can become very large (we
will show this below), the field theories of interest, i.e.,
finite density quantum field theory in the strong sign-problem regime,
should give advanced moments within bounds. In this case, the
convergence is then left to the coefficients $\alpha _n$. Inserting
\Eq(\ref{eq:k8}) into \Eq(\ref{eq:k5}), 
we find after a renaming of indices
\bea 
&M_{2N +2}& \nn\\ 
&=& \frac{ 1 }{k_{N} ^{(N)}}
\Biggl[  \sum_{n=0}^{N} \, k_n^{(N)} \, \la s^{2n+2} \ra 
- \sum _{k=2}^{N} \alpha _{2k} \, \sum _{n=1}^k \gamma _{kn} \la
s^{2n} \ra  \; \Biggr]
\nonumber \\
\\
&=& \la s^{2N+2} \ra 
\nonumber \\ 
&+&  \sum_{n=1}^N
\frac{ 1 }{k_{N} ^{(N)} } \Bigl( k_{n-1} ^{(N)} \, - \, 
\sum _{k=\mathrm{max}(n,2) }^{N} \alpha _{2k} \, \gamma _{kn} \Bigr) 
\, \la s^{2n} \ra  \;,
\nonumber
\\
\ena 
where we have changed the order of the double sum. We therefore
find the recursion: 
\bea 
\gamma _{N+1 \, n} &=& \Bigl( k_{n-1} ^{(N)} \, - \, 
\sum _{k=\mathrm{max}(n,2) }^{N}  \alpha _{2k} 
  \, \gamma _{kn} \Bigr)  / k_{N} ^{(N)}  
\\
\gamma _{N+1 \, N+1} &=& 1 \; , 
\label{eq:k9} 
\ena
where $ 1 \leq n \leq N$ and $2 \leq N \leq N_0 -1 $. 
The recursion can be solved in closed form for $i\in\{2,\ldots,N_0\} $
and $j\in\{1,\ldots,N_0\} $: 
\bea
\gamma_{ii} &=& 1 \;, \; \; \; \; \; 
\gamma_{21} \; = \;  \frac{k_0^{(1)}}{k_1^{(1)}}  \;, \; \; \; \; \; 
\gamma_{ij} = 0 \; \mbox{for}  \; j>i \; , 
\\
\gamma_{ij} &=& \frac{k_{j-1}^{(i-1)} - k_{j-1}^{(i-2)}}{k_{i-1}^{(i-1)}} \;,
\hbo i>j \mbox{ and }  i>2 \;. 
\label{eq:k99}
\ena
 
\subsection{The first advanced moments } 

For illustration purposes, we will explicitly calculate the first few
advanced moments. The main task is to obtain the coefficients
$k^{(N)}_i$, which emerge from the solution of a linear set of
equations, see \Eq(\ref{eq:k2b})).

\medskip 
For the leading order $N_0=2$, we find 
\be 
(A_{ij}) =  2 \, 
\begin{pmatrix}
  \ds \frac{\pi^3}{3} &\ds \frac{\pi^5}{5}  \\ \\ 
  \ds \frac{\pi^5}{5} &\ds \frac{\pi^7}{7}  
\end{pmatrix} , \hbo 
(I_{2j}) = 
\begin{pmatrix}
0 \\ \\ - 4 \pi 
\end{pmatrix}.
\en
\be
k^{(1)}_1 \, = \, \alpha _4 \, = \, - \frac{175}{2 \pi^6} \; , \hbo 
k^{(1)}_0/k^{(1)}_1 \, = \, - \frac{3}{5} \pi ^2 \; . 
\en
Hence, the first advanced moment, see \Eq(\ref{eq:k3}), is given by:
\be
M_4 \; = \; \la s^4 \ra \; - \; \frac{3}{5} \pi ^2 \, \la s^2 \ra \;
. 
\en
At next to leading order, i.e., $N_0=3$, we have 
\be
(A_{ij}) =  2 \, 
\begin{pmatrix}
  \ds \frac{\pi^3}{3} &\ds \frac{\pi^5}{5} &\ds \frac{\pi^7}{7}  \\ \\ 
  \ds \frac{\pi^5}{5} &\ds \frac{\pi^7}{7}   &\ds \frac{\pi^9}{9}  \\ \\
  \ds \frac{\pi^7}{7} &\ds \frac{\pi^9}{9}   &\ds \frac{\pi^{11}}{11}  
\end{pmatrix} , \; \; \; 
(I_{2j}) = 
\begin{pmatrix}
0 \\ \\ - 4 \pi \\ \\ -8 \pi ^3 + 58 \pi 
\end{pmatrix} .
\en
The solution of the corresponding linear system is given by 
\bea 
k^{(2)}_0 &=& - \frac{945}{8} \, \frac{ 2 \pi^2 - 33}{\pi^6} \; , 
\hbo 
\\  
k^{(2)}_1 &=& \frac{315}{4} \, \frac{ 16 \pi^2 - 231}{\pi^8} \; , 
\\
k^{(2)}_2 &=&  - \frac{4851}{8} \, \frac{ 2 \pi^2 - 27}{\pi^{10} } \;
= \; \alpha _6 \; . 
\ena 
From \Eq(\ref{eq:k9}), we then find for the coefficients $\gamma $ 
\bea 
\gamma _{31} &=& \frac{ k^{(2)}_0 - \alpha_4 \gamma_{21} }{k_2^{(2)}}
=  \frac{ 5}{21} \, \pi ^4 \; , 
\nonumber \\ 
\gamma _{32} &=& \frac{k^{(2)}_1 - \alpha_4 \gamma_{22} }{k_2^{(2)}} = 
- \frac{10}{9} \pi^2 \; , 
\nonumber \\ 
\gamma _{33} &=& 1 \; , 
\nonumber 
\ena
leaving us with: 
\be 
M_6 = 
\langle s^6 \rangle \; - \; \frac{ 10 \pi^2 }{9} \, \langle s^4 \rangle \;
+ \; \frac{ 5\pi^4 }{21} \, \langle s^2 \rangle \; . 
\label{eq:m6}
\en 
Up to order $N_0=3$, the phase factor expectation value is given by:
\bea
\la  \e^{i\phi} \ra_{PQ}  &=& - \frac{175}{2 \pi^6} \, M_4 
\; - \;  \frac{4851}{8} \, \frac{ 2 \pi^2 - 27}{\pi^{10} } \, M_6 \; . 
\nonumber 
\ena 
We have computed the moment coefficients up to order $N_0=5$. We find 
for the coefficient matrix $(k\ge 2, i \ge1)$: 

\be
(\gamma _{ki}) = 
\left(
\begin{array}{ccccc}
  \ds -\frac{3 \pi ^2}{5}    &   1                      & 0                     & 0                     & 0  \\
  \\
  \ds \frac{5 \pi ^4}{21}    &\ds   -\frac{10 \pi ^2}{9}   & 1                     & 0                     & 0  \\
  \\
  \ds -\frac{35 \pi ^6}{429} &\ds   \frac{105 \pi ^4}{143} &\ds -\frac{21 \pi ^2}{13} & 1                     & 0  \\
  \\
  \ds \frac{63 \pi ^8}{2431} &\ds   -\frac{84 \pi ^6}{221} &\ds \frac{126 \pi ^4}{85} &\ds -\frac{36 \pi ^2}{17} & 1  \\
  \\
\end{array}
\right) 
\label{eq:gamma} 
\en
and for the lead coefficient in front of the advanced moments:

\medskip 
{\small 
\begin{tabular}{ll}
$\alpha _4$ & $\ds = -\frac{175}{2 \pi ^6} $ \\ & \\ 
$\alpha _6$ & $ \ds =-\frac{4851 \left(-27+2 \pi ^2\right)}{8 \pi ^{10}}
              $ \\& \\ 
$\alpha _8$ & $\ds =-\frac{57915 \left(2145-242\pi ^2+3 \pi
              ^4\right)}{16 \pi ^{14}} $ \\ & \\ 
$\alpha _{10}$ & $\ds =-\frac{2540395 \left(-348075+44850 \pi ^2-1014
                 \pi ^4+4 \pi ^6\right)}{128 \pi ^{18}} $ 
\end{tabular} 
}

\medskip
We finally perform a consistency check. For a truncation of the
density-of-states at order $N_0$, 
all the moments up to  $M_{2N_0}$  contribute to the 
the phase
factor expectation value at this order, see \Eq(\ref{eq:k7}).
If we consider \Eq(\ref{eq:expansion})
as exact for the moment in the sense that all simple moments $\la s^{2n} \ra $
are calculated with this density, then the phase
factor expectation value is obtained exactly by summing all
contributions including the term containing $M_{2N_0}$. Since this
result is already exact, all moments $M_{2k}$ with $k> N_0$ must
vanish. For example, assume that the density is given by 
$$ 
\rho (s) \; = \; d_0 + d_1 s^2 \;, \hbo (N_0=2) , 
$$
then e.g.~$M_6$ (and all higher moments need to vanish for all choices
for $d_0$ and $d_1$. This devises a consistency check. We find for the
present example: 
\bea 
\la s^6 \ra &=& \frac{ 2 \pi ^7 }{63} \, \Bigl(9 d_0 + 7 \pi ^2 d_1
\Bigr) \; , 
\nonumber \\
\la s^4 \ra &=& \frac{ 2 \pi ^5 }{35} \, \Bigl(7 d_0 + 5 \pi ^2 d_1
\Bigr) \; , 
\nonumber \\
\la s^2 \ra &=& \frac{ 2 \pi ^3 }{15} \, \Bigl(5 d_0 + 3 \pi ^2 d_1
\Bigr) \; . 
\nonumber 
\ena 
Inserting these simple moments into $M_6$, \Eq(\ref{eq:m6}), we find that
all terms cancel and that $M_6$ indeed vanishes for all choices of
$d_0$ and $d_1$. If we consider, in a quantum field theory setting,
the expansion \Eq(\ref{eq:expansion}) as an expansion with respect to
some inverse power of the volume, the moments $M_{2n}$ are then 
suppressed by these powers. 

\subsection{Convergence} 

\begin{figure}[t]
  \begin{center}
    \includegraphics[width=0.4\textwidth]{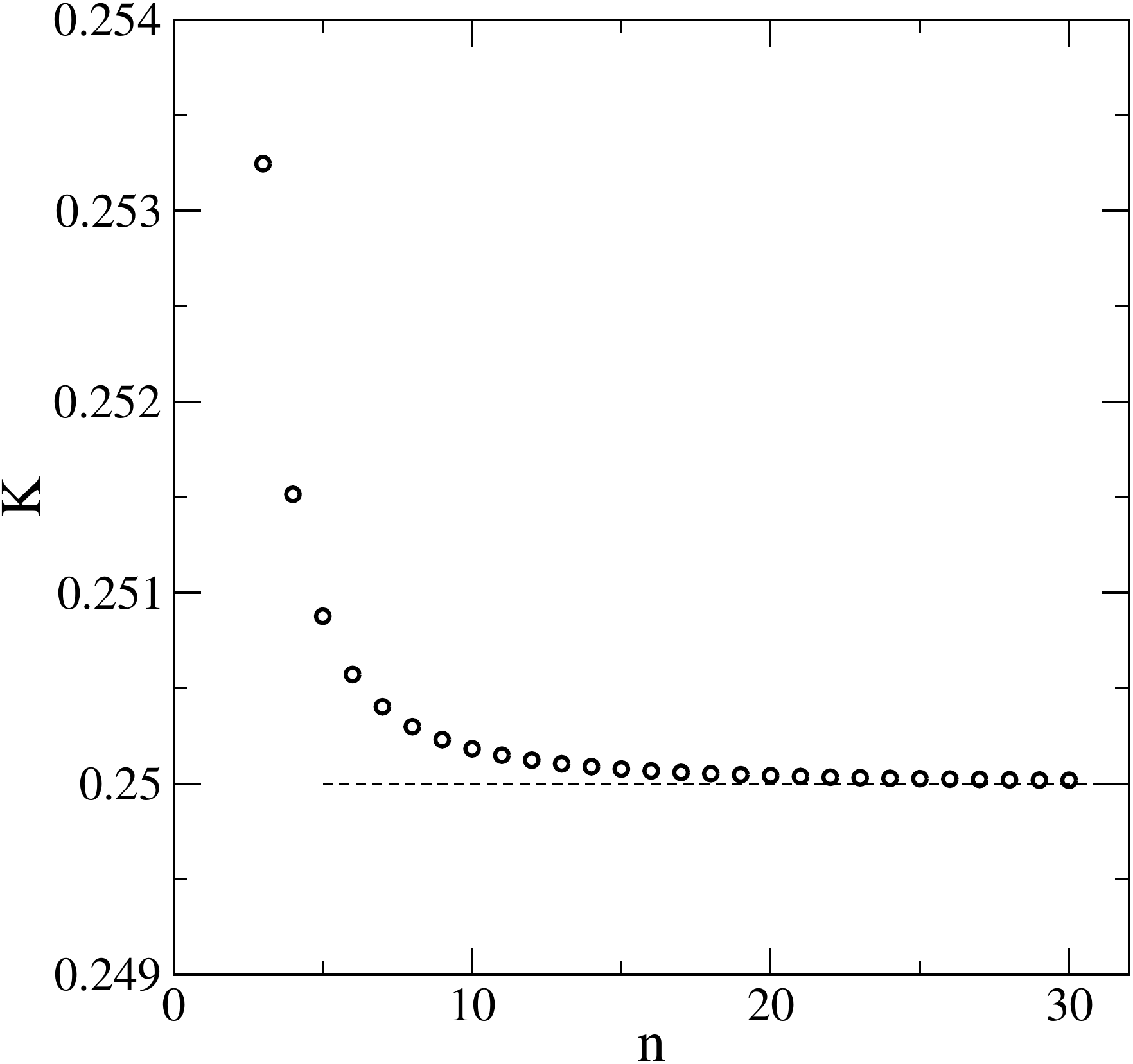}
  \caption{The exponential rate $K$, see (\ref{eq:k15}), as a function of
    the order $n$ of the expansion.}
  \label{fig:L} 
    \end{center}
\end{figure}
\begin{figure}[t]
  \begin{center}
    \includegraphics[width=0.4\textwidth]{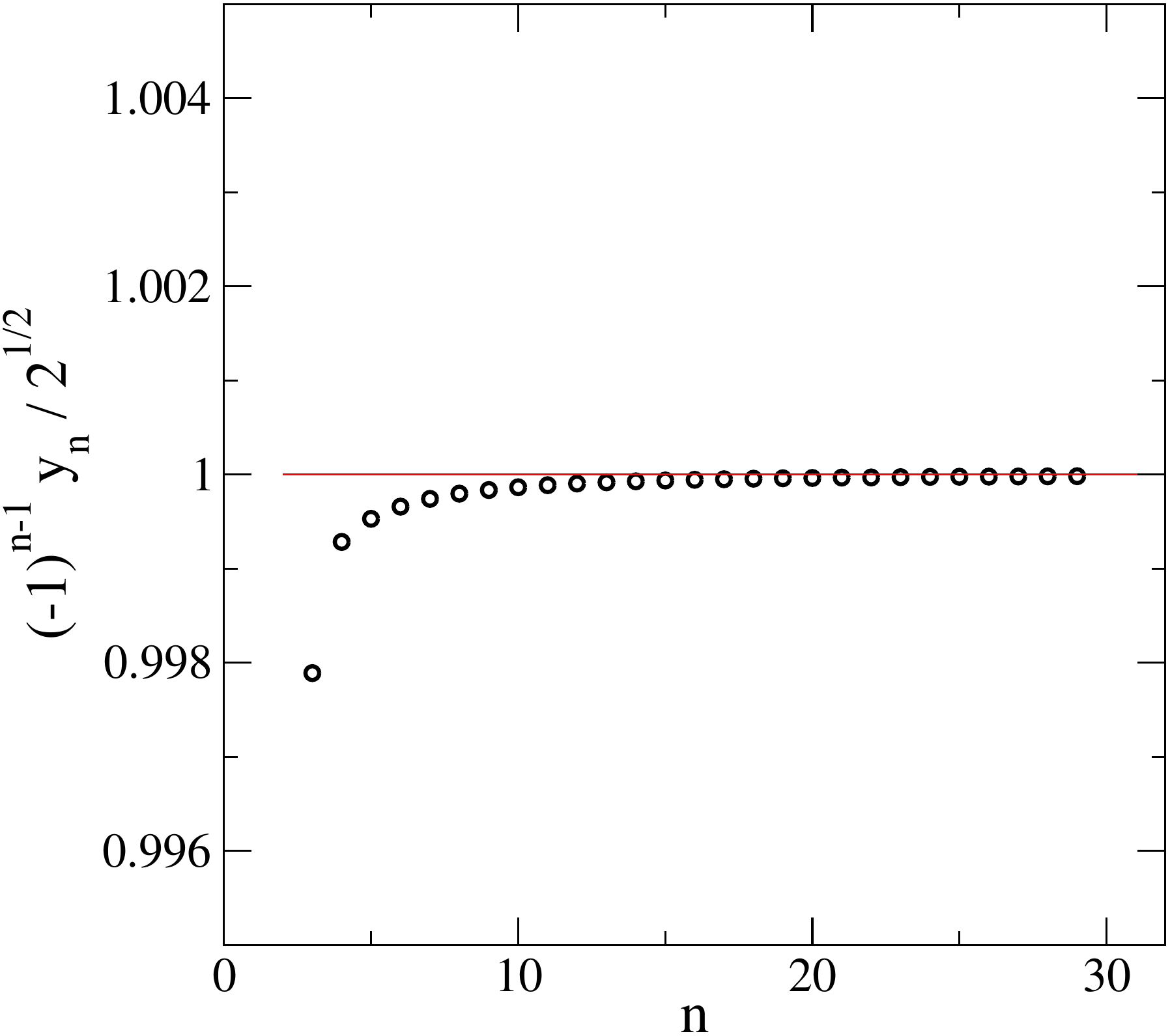}
  \caption{ The highest element $y_n$ as a function of  the order $n$ of the expansion.}
  \label{fig:y} 
    \end{center}
\end{figure}
\begin{figure}[t]
  \begin{center}
    \includegraphics[width=0.4\textwidth]{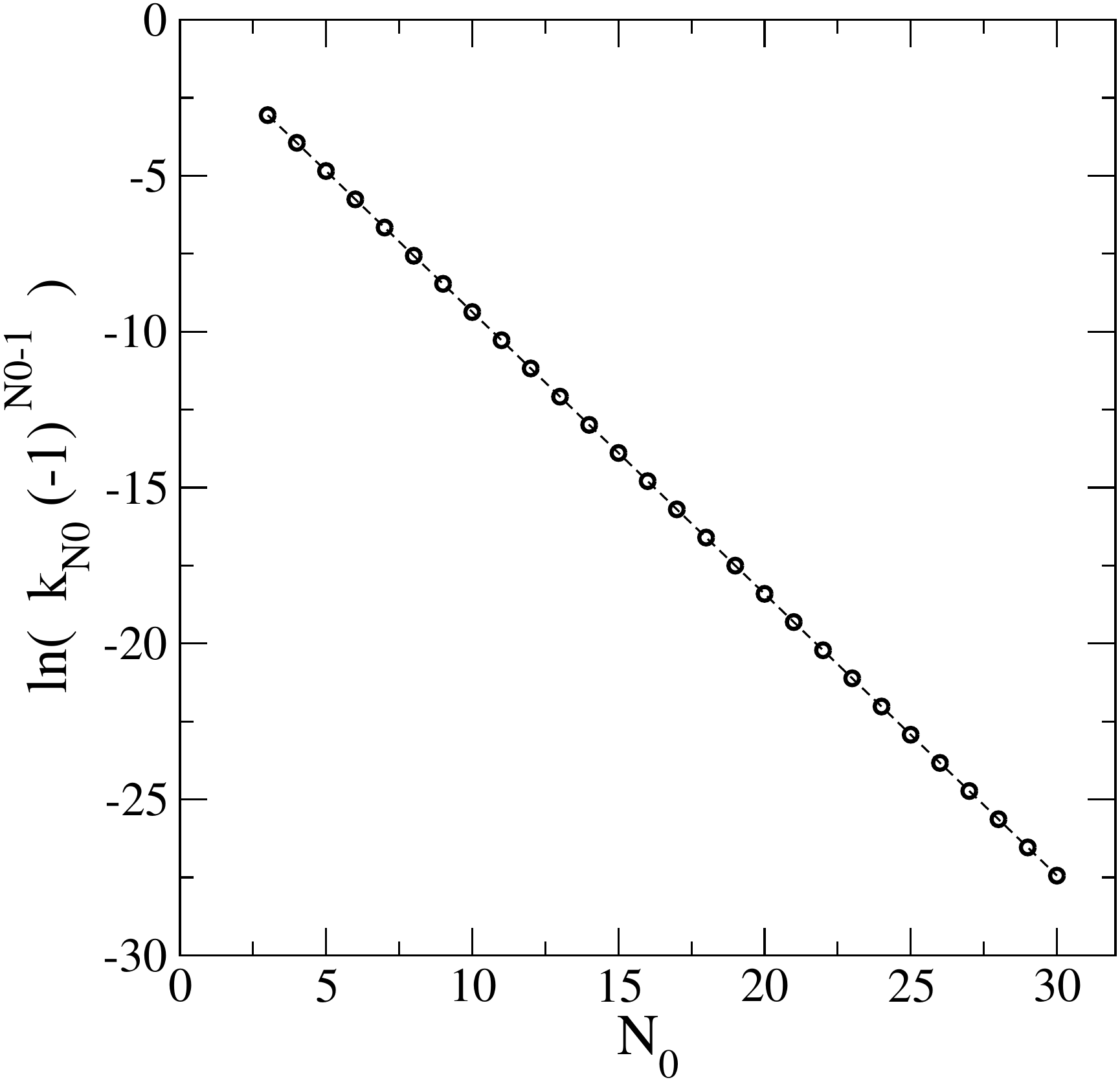}
  \caption{ The behaviour of the coefficient $\alpha _{2N_0}=k^{N_0}_{N_0}$
    of the expansion in terms of advanced moments.  }
  \label{fig:k} 
    \end{center}
\end{figure}
For high orders $N_0$, the coefficients $\gamma $ in the definition
\Eq(\ref{eq:k8}) of the advanced moments $M_{2k}$ can become very
large. In this section, we will assume that for functions $\rho(s)$
arising in a quantum field theory setting the moments remain
within bounds. This occurs due to cancellations between simple moments
$\la s^{2i} \ra $, as we will show below.
In this case, the expansion \Eq(\ref{eq:k7}) of the
phase factor expectation value in terms of the advanced moments is
dictated by behaviour of the coefficients $\alpha _{2k}$ for large
$k$. These coefficients are universal: they do not depend on the
underlying theory, i.e., $\rho (s)$. They arise from the solution of
the linear system  \Eq(\ref{eq:k2b}), which reads in a shorthand notation 
\be 
k \; = \; A^{-1} \, I \; , 
\en
and it is this linear system that we are going to
study in greater detail. 
Since the matrix $A$ in \Eq(\ref{eq:Aij}) is symmetric and positive, 
we perform a Cholesky decomposition and solve for $k$: 
\be
A = L L^T, \hbo L \, y \, = \, I , \hbo L^T k=y \; , 
\en
where $L$ is a lower triangular matrix. 
Note that if the system  $Ly=b$ is solved at order $N_0$ and if
subsequently the order $N_0$ is increased, the first $N_0$ components
of the solution $y$ are unaffected by the increase due to the
triangular form of $L$. The same is true for the matrix $L$:
increasing the order from $N_0$ to $N_0+1$
does not affect the first $N_0$ rows and columns.
We are interested in the $N_0$ dependence of the 
last component of $k$: 
\be 
k^{(N_0)}_{N_0} = y_{N_0} / L_{N_0,N_0}  \; . 
\label{eq:11}
\en 
The Cholesky decomposition gives
\bea
L_{ii} &=& \sqrt{ A_{ii} - \sum_{k=1}^{i-1}  L_{ik}^2 } \;, 
\label{eq:k10} \\
L_{ij} &=& \frac{1}{L_{ij}} (A_{ij} - \sum_{k=1}^{j-1}  L_{ik}L_{jk})
\;,\qquad i>j 
\nonumber 
\ena
We have solved this iteration analytically for values $N_0$ up to
$30$. We find that for large $n$ the data is well described by 
We find that very quickly $L_{nn}$ reaches an asymptotic regime
which is well describe by
\be
L_{n+1,n+1} \; = \; K \, \pi ^2 \, L_{nn} \; , \hbo K=\frac{1}{4}\;,
\label{eq:k15} 
\en
see \Fig{\ref{fig:L}}. Asymptotically, we therefore find the
exponential increase: 
\be 
L_{N_0, N_0} \; \propto \; \left( \frac{1}{4} \, \pi^2 \,
\right)^{N_0} \; . 
\label{eq:16}
\en 
In a next step, we studied the asymptotic behaviour of the solution
$y$ of the linear system $L \, y \, = \, I$. 
We find numerical evidence (see \Fig{\ref{fig:y}}) that
$y_n$ converges quickly to a constant
\be 
\lim _{n \to \infty } y_n \; = \; \sqrt{2} \; . 
\label{eq:17}
\en 
This suggest that the asymptotic $N_0$ dependence of the desired
expansion coefficient is given by: 
\be 
\alpha _{N_0} \; \propto \; \left( \frac{1}{4} \, \pi^2 \,
\right)^{ - \, N_0} \; . 
\label{eq:18}
\en 
Unfortunately, we could not prove any of these asymptotic behaviours
analytically, but we have verified \Eq(\ref{eq:18}) by also solving the
linear system $L^T k = y $ for $k$. Our analytical result for $N_0=2$
to $N_0=32$ is shown in \Fig{\ref{fig:k}}. We find the remarkable
result that the expansion coefficients $\alpha _{2N_0}$ are
exponentially decreasing with $N_0$ suggesting a rapid convergence of
the Advanced Moment expansion as long as the moments $M_{2n}$ are
bounded.
    

\subsection{Application to HDQCD}

In essence, the Advanced Moments approach from section~\ref{sec:am}
is an efficient numerical method to evaluate the Fourier transform
\Eq(\ref{eq:phase_rhoF})
for sufficiently smooth integrands $\rho (s)$. In this section, we
test the method in the quantum field theory context of QCD at finite
densities of heavy quarks (HDQCD).
Our preliminary results have been reported in~\cite{Garron:2016nrm}.

Here we are interested in the strong sign problem region (in which
$\mu \sim 1.3$): in \Fig{\ref{fig:3}}, we show that the density
is almost constant 
whereas for $\mu\sim 1$ and $\mu\sim 1.4$, the density has
variation of order $1$ (see \Figs{\ref{fig:1}} and \ref{fig:2}).
Hence, we expect that the Advanced Moment expansion will have a better
convergence in the strong sign problem regime.

From now on, we focus on the severe sign problem region,  $\mu=1.2921$.
Once the density is known, we can compute the
elementary moments (again using our fit results and semi-analytic integration).
They are reported in \Table{\ref{table:moments}}.
By virtue of the LLR method, they are extracted with a very good
statistical precision. 
\begin{table}[t]
  \begin{tabular}{cccc}
    \hline \\[-2ex]
    Moment & Central Value & Error & Rel. Error ($\%$) \\[1ex]
    \hline \\[-2ex]
    $\la s^2 \ra$   & $\z\z\z3.\,289\,859\, 4$ & $\z\z  13 \times 10^{-7}$  & $3.9 \times 10^{-5}$\\
    $\la s^4 \ra$   & $\z\z 19.\,481\,750\, 1$ & $\z   100 \times 10^{-7}$  & $5.1 \times 10^{-5}$\\ 
    $\la s^6 \ra$   & $\z  137.\,340\,787\, 5$ & $\z   778 \times 10^{-7}$  & $5.7 \times 10^{-5}$\\ 
    $\la s^8 \ra$   & $   1054.\,276\,996\, 8$ & $    6251 \times 10^{-7}$  & $5.9 \times 10^{-5}$\\
    $\la s^{10}\ra$ & $   8513.\,423\,834\, 6$ & $   51793 \times 10^{-7}$  & $6.1 \times 10^{-5}$\\
    \hline
  \end{tabular}
  \caption{First elementary moments for $\mu=1.2921$}
  \label{table:moments}
\end{table}
We also observe that going from $\la s^2\ra$ to $\la s^8\ra$, the
relative error increases very slowly. We turn now to the advanced moments:
since all the elementary moments are positive, the relative signs in
(\ref{eq:M4})-(\ref{eq:M8}) imply that important cancellations occur.
At leading order (LO), we have
\bea
M_4 &=& \langle s^4 \rangle - \frac{ 3 \pi^2 }{5} \, \langle s^2 \rangle \; , \\
&=& 19.\,481\,750\,4(100) - 19.\,481\,766\,3 (77)  \;,\\
\label{eq:M4}
&=& -0.\,000\,015\,9(23) \;,
\ena
 and at next-to-leading order (NLO) we find: 
\bea
M_6 &=& 
\langle s^6 \rangle \; - \; \frac{ 10 \pi^2 }{9} \, \langle s^4 \rangle \;
+ \; \frac{ 5\pi^4 }{21} \, \langle s^2 \rangle \; , \\
&=&
137.\,340\,787\, (78) -213.\,641\,300\, (110) \nn \\
&&  + 76.\,300\,527\, (30)    
\;,\\
&=&
\label{eq:M6}
0.\,000\,014\, 2(21)   \;,
\ena
%
where for next-to-next-to leading order (NNLO), we obtain: 
\bea
M_8 &=&  \langle s^8 \rangle \; - \; \frac{ 21 \pi^2 }{13} \, \langle
s^6 \rangle \; 
\nn\\
&&
+ \; \frac{ 105\pi^4 }{143} \, \langle s^4 \rangle \; 
- \; \frac{ 35\pi^6 }{429} \, \langle s^2 \rangle \;, \\
&=&
 1054.\,277\,0\, (6)
-2189.\,652\,6\, (12) \nn\\
&&
+1393.\,415\,8\, (7)
-258.\,040\,2\, (1)
\;,\\
&=&
\label{eq:M8}
-0.\,000\,003\, 5 \,(5)   \;.
\ena
%
As expected, strong cancellations between the simple moments occur
making it mandatory to determine the simple moments with high precision.
The analysis has been carried out using the bootstrap resampling method,
and we point out that strong correlations are at work to obtain the
Advanced Moments at the level of precision reported here.
The numerical values are also
reported in \Table{\ref{table:advmoments}}. One should note that the
overall sign of the advanced moments oscillate, however $\alpha_i M_i$
is a positive quantity, as can be seen in \Eq(\ref{eq:k7}), or in the
numerical values. 
\\
\begin{table}[t]
  \begin{tabular}{cccc}
    \hline \\[-2ex]
    Moment & Central Value & Error & Rel. Error ($\%$) \\[1ex]
    \hline \\[-2ex]
    $M_4$    & $ -1.592\times 10^{-5}$ & $2.35\times 10^{-6}$ & $15\%$ \\
    $M_6$    & $\m1.424\times 10^{-5}$ & $2.11\times 10^{-6}$ & $15\%$ \\
    $M_8$    & $ -3.503\times 10^{-6}$ & $5.18\times 10^{-7}$ & $15\%$ \\
    $M_{10}$ & $\m4.205\times 10^{-7}$ & $6.22\times 10^{-7}$ & $15\%$ \\
    \hline
  \end{tabular}
  \caption{First advanced moments for $\mu=1.2921$. Notes that the
    relative signs   cancel out with those of the coefficients
    $\alpha_i$.} 
  \label{table:advmoments}
\end{table}
\\

The phase factor expectation value \Eq(\ref{eq:k7}) is then given by
\bea
\la \e^{i\phi} \ra
= 10^{-6} \times &\big(&
     1.45(21) \hbo \,\;\mbox{LO} \nn \\
&+&  0.67(10) \hbo \,\; \mbox{NLO} \nn \\
&+&  0.068(10) \hbo \mbox{NNLO} \nn \\
\label{eq:conv1}
     &+&  \ldots  \big) \;, \\
\label{eq:conv2}
= 2.186 (323) \times 10^{-6} &+& {\cal O}(\alpha_{5}M_{5}) \;.
\ena
When the order of the expansion increases, the statistical error decreases
and that the results converges quickly to the ``exact'' answer     
\be
\la \e^{i\phi} \ra = 2.189(324) \times 10^{-6}\;,
\en
obtained by fitting the
extensive density $\rho _E$ and by carrying out the
Fourier transform using the fit, as in~\cite{Garron:2016noc}.
(In the latter we quote $2.37(21) \times 10^{-6}$, the small difference
in the central value comes from the fact that we use a different $\delta_s$).
We observe a rapid convergence here. 

Since the phase factor is a small number, it is useful to look at the
logarithm of this quantity. We find 
\be
\log \la \e^{i \phi} \ra_{PQ}
= -13.032  \pm 0.152 
\qquad (\rm Full) \;.
\en
The Advanced Moment method yields 
\bea
&& \log \la \e^{i \phi} \ra_{PQ} \\
&=& -13.445 \pm 0.152 + {\cal O}(\alpha_{6}  M_6) \qquad \;\;(\rm LO) \;,\\
&=& -13.065 \pm 0.152 + {\cal O}(\alpha_{8}  M_8) \qquad \;\; (\rm NLO) \;,\\
&=& -13.033 \pm 0.152 + {\cal O}(\alpha_{10}  M_{10}) \qquad (\rm NNLO).
\ena
It is remarkable that not only the central value but also the variance is very
well approximated by our expansions. Indeed for this value of $\mu$,
the full (relative) variance is already given by the first order.
Of course the quality of the approximation depends
on the variation of $\rho $ (and therefore on the strength of the sign
problem).

\begin{figure}[t]
  \begin{center}
  \includegraphics[width=0.5\textwidth]{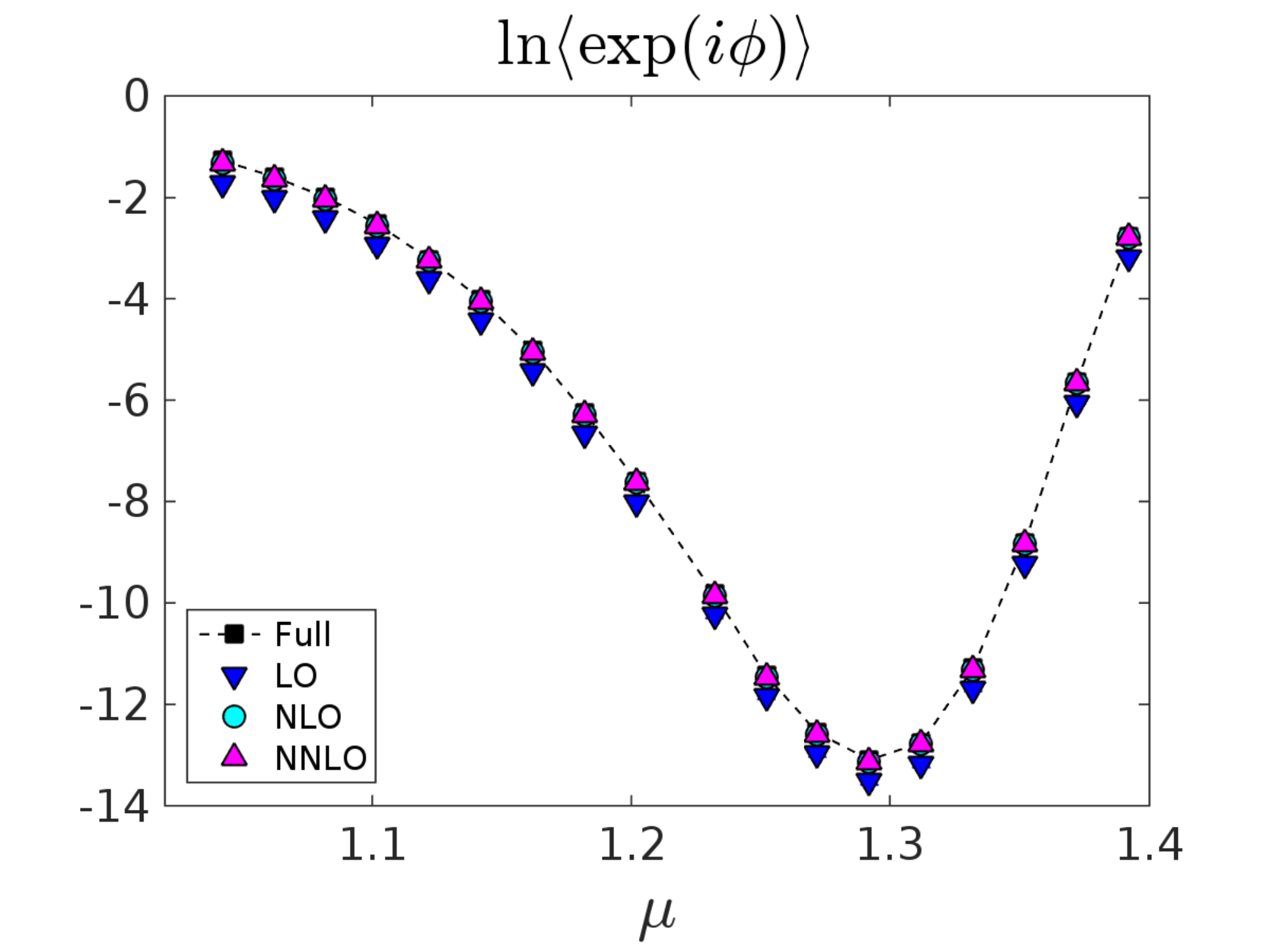}
  \caption{Comparison of the phase factor expectation value,
    computed with the ``full theory'', with the result from the
    Advanced Moments method.
    We observe that the moment expansion rapidly converges with 
    the NNLO and NNLO lie on top of each other and are indistinguishable
    from the full answer. Statistical error bars are included.}
  \label{fig:9}
  \end{center}
\end{figure}

\begin{figure}[t]
  \begin{center}
  \includegraphics[width=0.5\textwidth]{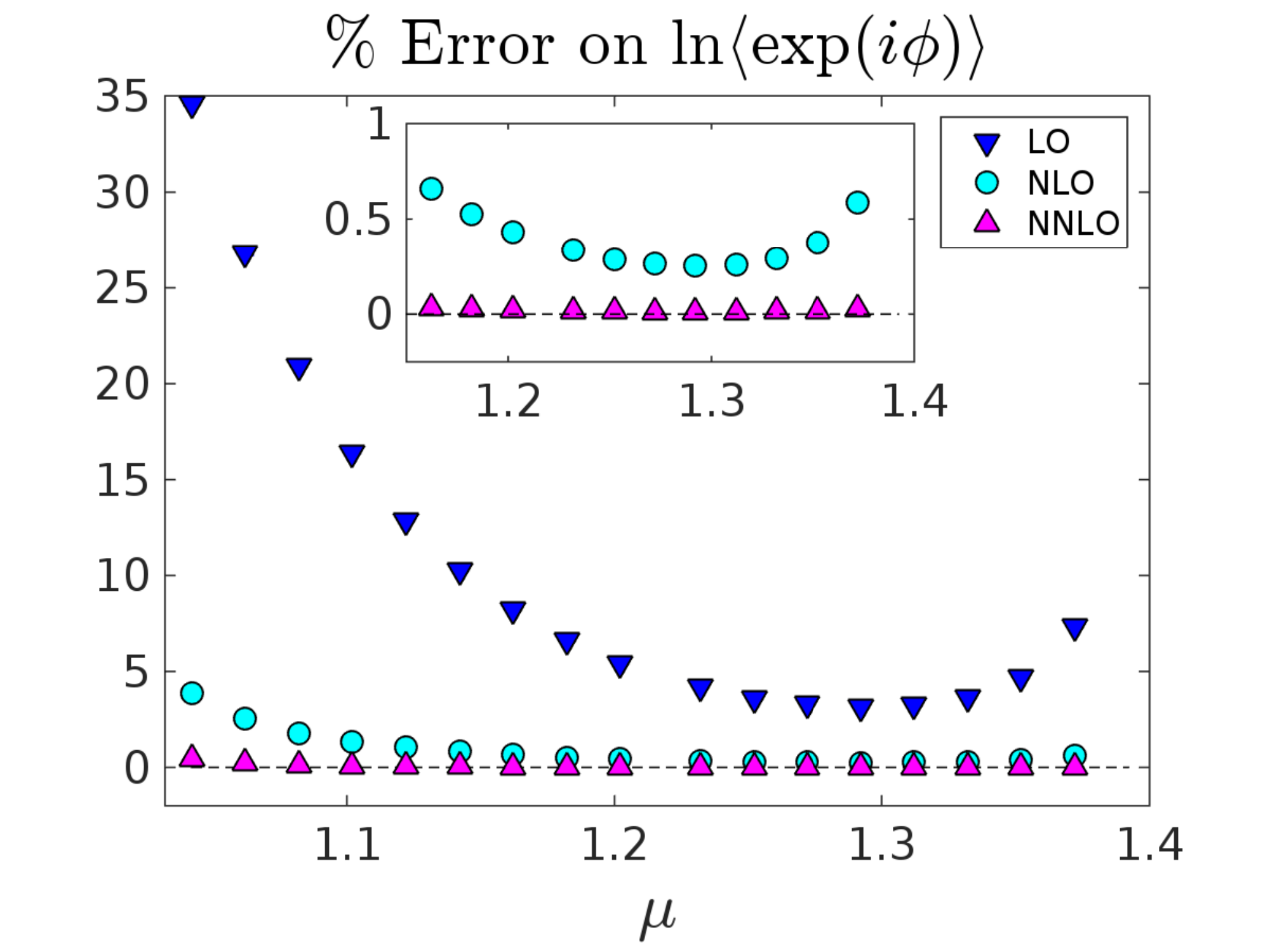}
  \caption{Relative difference between the moment approximation and
    the full answer. As expected, the approximation works better
    in the strong sign problem regime} 
  \label{fig:10}
  \end{center}
\end{figure}

We now vary the value of $\mu$ in the range $1<\mu<1.4$
and compare the results of the phase factor expectation value obtained
in~\cite{Garron:2016noc}
with the method proposed here.
It is interesting to note that even in the weak sign-problem region,
in which the density $\rho $ fluctuates between $0$ and $1$,
the NLO and NNLO approximations already yield decent approximations. 
This is illustrated in \Figs{\ref{fig:9}} and \ref{fig:10}.
Our numerical results can be found in \Table{\ref{table:resultsmu}}.
We quote the ``full answer'' as obtained in~\cite{Garron:2016noc}
and the relative difference with the method presented here,
for the first three orders.
(Here we implement the Advanced Moments method with the same $\delta_s$
as in~\cite{Garron:2016noc}.)
The NLO approximation
works at the percent level over the full available range,
even in the weak-sign problem region.

\begin{table}[htb]
   \begin{tabular}{c| c c c c  }
     \hline \\[-2ex]
     $\mu$ & $\ln \la \exp(i\phi)\ra$ & LO ($\%$) & NLO ($\%$) & NNLO ($\%$) \\
     \hline
 1.042 & -1.271(2)  &  36  &  4.1  &  0.48 \\
 1.062 & -1.588(3)  &  27  &  2.6  &  0.22 \\
 1.082 & -1.993(3)  &  21  &  1.8  &  0.11 \\
 1.102 & -2.52(0)   &  16  &  1.3  &  0.07 \\
 1.122 & -3.213(5)  &  13  &  1.0  &  0.05 \\
 1.142 & -4.019(10) &  10  &  0.8  &  0.04 \\
 1.162 & -5.02(1)   &  8   &  0.7  &  0.03 \\
 1.182 & -6.251(10) &  7   &  0.5  &  0.02 \\
 1.202 & -7.602(25) &  5   &  0.4  &  0.02 \\
 1.232 & -9.823(66) &  4   &  0.3  &  0.02 \\
 1.252 & -11.43(5)  &  4   &  0.3  &  0.01 \\
 1.272 & -12.56(6)  &  3   &  0.3  &  0.01 \\
 1.292 & -13.0(2)   &  3   &  0.3  &  0.01 \\
 1.312 & -12.76(7)  &  3   &  0.3  &  0.01 \\
 1.332 & -11.29(4)  &  4   &  0.3  &  0.01 \\
 1.352 & -8.811(13) &  5   &  0.4  &  0.02 \\
 1.372 & -5.639(20) &  7   &  0.6  &  0.03 \\
 1.392 & -2.756(8)  &  15  &  1.2  &  0.06
   \end{tabular}
   \caption{Logarithm of the phase factor expectation value from ~\cite{Garron:2016noc}
    and relative precision obtained with the moment method
    for the various orders as a function of $\mu$.}
      \label{table:resultsmu}
 \end{table}

\section{Conclusions}
\label{sec:conclusions}

There are two main possibilities in addressing finite density quantum
field theory: (i) facing the large cancellations that give rise to the
smallness of the partition function or (ii) to reformulate to an
equivalent theory say by dualisation~\cite{Gattringer:2014nxa} or by a
complexfication of the fields~\cite{Aarts:2013lcm}. Method (ii) would
be preferred if the approach exists and if exactness can be
guaranteed. The appeal of method (i) is that it is universally
applicable if a way is found to control the cancellations.

\medskip
A first success for direction (i) emerged with the advent of
Wang-Landau type techniques and, most notably, the LLR
method~\cite{Langfeld:2012ah}: due to the feature of exponential error
suppression of the LLR approach~\cite{Langfeld:2015fua}, high
precision data for the density-of-states $\rho (s)$ of finding a
particular phase $s$ over many orders of magnitude has become
available. The partition function now emerges as Fourier transform of
$\rho (s)$. Due to large cancellations, this Fourier transform is a
challenge in its own right. The recent success reported
in~\cite{Langfeld:2014nta} and in~\cite{Garron:2016noc} hinge on the
ability to find a fit function for $\ln \rho (s)$ that well represents
hundreds of numerical data points with relatively few fit
parameters. This situation is unsatisfactory since the quest for this
fit function might not be always successful.

\medskip
The present paper explores three methods to perform the Fourier
transform:

\begin{itemize}
  \item[$\bullet$] The {\it Gaussian } approximation of the extensive
    density-of-states $\rho _E$ is most easily implemented, but hard
    to improve in a systematic way. For the example of HDQCD, we found
    this approximation yields the right order of magnitude through out
    and only misses the exact phase factor by a factor of two when the
    sign problem is strongest.

  \item[$\bullet$] The {\it telegraphic } approximation yields the phase factor
    through an alternating (discrete) sum of the extensive
    density-of-states $\rho _E$. The relative systematic error is of
    the order of the phase factor itself, which makes the
    approximation excellent in the strong sign-problem regime.

  \item[$\bullet$] The {\it advanced moment} approach is a systematic
expansion of this Fourier transform with respect to the deviations of
$\rho (s)$ from uniformity. The expansion therefore works best in the
strong-sign problem regime. The expansion is independent of the
quantum field theory setting an can be applied to the Fourier transform
of any sufficiently smooth function $\rho (s)$, $s \in [-\pi,\pi]$. At
the heart of expansion are the so-called Advanced Moments. We have
thoroughly derived these moments and the theory independent expansion
coefficients $\alpha $. We found evidence that
the expansion coefficients decrease exponentially with increasing order, thus
guaranteeing rapid convergence if $\rho (s)$ admits
moments $M$ that are bounded.
We have tested and validated the Advanced Moment
expansion in the context of HDQCD: we have confirmed that the
expansion converges very quickly. It works best in the strong-sign
problem region as expected, although at third order the results
agree with the ``full'' answer at the sub-percent level
even in the weak sign problem regime.
  
\end{itemize}

{\bf Acknowledgements: } 
We are grateful to B.~Lucini and A.~Rago for helpful
discussions.
NG and KL are
supported by  the  Leverhulme Trust (grant RPG-2014-118) and, KL by
STFC (grant ST/L000350/1). 

\bibliography{mybibfile}

\end{document}